\documentclass[aps,prl,floatfix,twocolumn]{revtex4}

\usepackage{graphics}
\usepackage{epsfig}

\begin{document}

\title{Second Renormalization of Tensor-Network States}

\author{Z. Y. Xie$^1$}
\author{H. C. Jiang$^2$}
\author{Q. N. Chen$^1$}
\author{Z. Y. Weng$^2$}
\author{T. Xiang$^{3,1}$}\email{txiang@aphy.iphy.ac.cn}

\affiliation{$^1$Institute of Theoretical Physics, Chinese Academy
of Sciences, P.O. Box 2735, Beijing 100190, China}

\affiliation{$^2$Center for Advanced Study, Tsinghua University,
Beijing, 100084, China}

\affiliation{$^3$Institute of Physics, Chinese Academy of Sciences,
P.O. Box 603, Beijing 100190, China}

\date{\today}

\begin{abstract}

We propose a second renormalization group method to handle the
tensor-network states or models. This method reduces dramatically
the truncation error of the tensor renormalization group. It allows
physical quantities of classical tensor-network models or
tensor-network ground states of quantum systems to be accurately and
efficiently determined.

\end{abstract}

\pacs{05.10.Cc,75.10.Jm,71.10.-w}

\maketitle

One of the biggest challenges in physics is to develop accurate and
efficient methods that can solve many currently intractable problems
in correlated quantum or statistical systems. While the density
matrix renormalizatoin group (DMRG) has proven to be a powerful
numerical tool for the study of strongly correlated systems in one
dimension, applications to two or higher dimensions are hampered by
accuracy. Quantum Monte Carlo simulations, on the other hand, are
not limited by the dimensionality, but are hamstrung by the minus
sign problem for fermionic or frustrated spin systems. To resolve
these difficulties, increasing interest has recently been devoted to
the study of the tensor-network states or
models\cite{Niggemann,Verstraete,Levin2007,Jiang2008,Gu}.

In statistical physics, all classical lattice models with local
interactions, such as the Ising model, can be written as
tensor-network models. To investigate these tensor-network models,
Levin and Nave proposed a tensor renormalization group (TRG)
method\cite{Levin2007}. They showed that the magnetization obtained
with this method for the Ising model on triangular lattice agrees
accurately with the exact result.

In a quantum system, a tensor-network
state\cite{Niggemann,Verstraete} presents a higher-dimensional
extension of the one-dimensional matrix-product
state\cite{Ostlund1995} in the study of DMRG\cite{White1992}. It
captures accurately the nature of short-range entanglement of a
quantum system and is believed to be a good approximation of the
ground state. In a recent work, we have developed a projection
method to determine accurately and systematically the tensor-network
ground state wavefunction for an interacting quantum
Hamiltonian\cite{Jiang2008}. In the evaluation of its expectation
values, we adopted the TRG method of Levin and Nave\cite{Levin2007}.
From the calculation, we found that the TRG can indeed produce
qualitatively correct results. However, the truncation error in the
TRG iteration grows rapidly with the bond dimension of local tensors
($D$). This leads to a big error in the calculation of expectation
values. In particular, the ground state energy and other physical
quantities oscillate strongly with increasing $D$, indicating that
the truncation error of the TRG is too big to produce a converging
result in the large $D$ limit.

In this Letter, we propose a novel renormalization group scheme to
solve the above problem. In the TRG method of Levin and Nave, the
singular-value spectra of an $M$-matrix defined by a product of two
neighboring local tensors is renormalized in the truncation of basis
space. This can be thought as the first renormalization to the
tensor-network state. However, this renormalization does not
consider the influence of other tensors (denoted as the environment
hereafter) to the $M$-matrix. It presents a local rather than global
optimization of the truncation space. The role of environment is to
modify the truncation space by reweighing the singular-value spectra
of $M$. We will introduce a systematical method to study this
renormalization effect of environment. This method, as will be
demonstrated below, improves significantly the accuracy of results.
We will call it the second renormalization group method of
tensor-network states, abbreviated as SRG.

\begin{figure}[tbp]
\includegraphics[width=0.9\columnwidth]{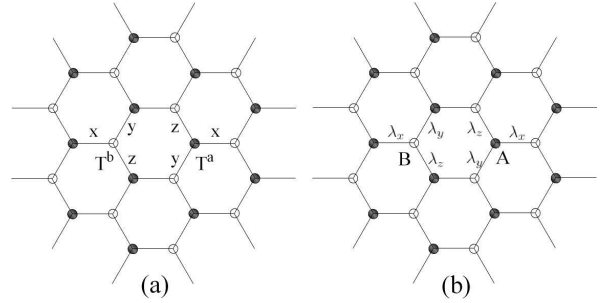}
\caption{Schematic representation of (a) the partition function of a
classical system defined by Eq.~(\ref{eq:z}) and (b) the
tensor-network ground state wavefunction of a quantum system defined
by Eq.~(\ref{eq:tns}) on honeycomb lattices. At each vertex, three
bonds are emitted along $x$, $y$, and $z$-directions, respectively.
} \label{fig:z}
\end{figure}

To understand how our method works, let us first consider how the
tensor-network state is renormalized in the TRG\cite{Levin2007}. We
start with a classical tensor-network model on honeycomb lattices
whose partition function is defined by
\begin{eqnarray}\label{eq:z}
Z = \mathrm{Tr} \prod_{i\in b, j\in w} T^a_{x_i y_i z_i} T^b_{x_j
y_j z_j},
\end{eqnarray}
where '$b/w$' stands for the black/white sublattice shown in
Fig~\ref{fig:z}. $T^a_{x_i y_i z_i} $ and $T^b_{x_j y_j z_j}$ are
the two tensors of rank three defined on the black and white
sublattices, respectively. The subscripts $x_i$, $y_i$, and $z_i$
are the integer bond indices of dimension $D$ defined on the three
bonds emitted from site $i$ along the $x$, $y$, and $z$ directions,
respectively. A bond links two sites. The two bond indices defined
from the two end points take the same values.

\begin{figure}[tbp]
\includegraphics[width=0.9\columnwidth]{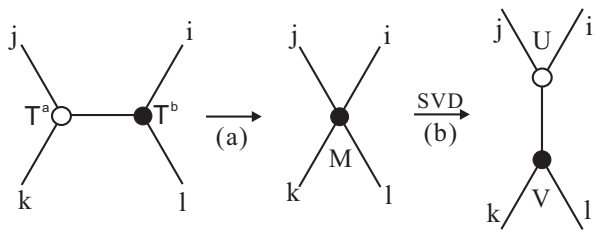}
\caption{(a) To form the $M$ matrix by tracing out the common bond
indices of tensors $T^a$ and $T^b$. (b) To perform the singular
value decomposition defined by Eq.~(\ref{eq:svd}). } \label{fig:sys}
\end{figure}

The TRG starts by rewiring a pair of tensors with singular value
decomposition as shown in Fig.~\ref{fig:sys}. To do this, let us
contract a pair of neighboring tensors to form a $N\times N$ matrix
$M$ defined by
\begin{equation}\label{eq:M}
M_{ij, kl} = \sum_{m} T^a_{mjk} T^b_{m li} ,
\end{equation}
where $N = D^2$. The singular value decomposition is then applied to
decouple this matrix into the following form
\begin{equation} \label{eq:svd}
\label{eq:T} M_{ij, kl} = \sum_{m= 1}^N U_{ij , m} \Lambda_m V_{kl ,
m},
\end{equation}
where $U$ and $V$ are two $N\times N$ unitary matrices. $\Lambda =
(\Lambda_1, \cdots , \Lambda_N)$ is a semi-positive diagonal matrix
arranged in descending order, $\Lambda_1 \ge \Lambda_2 \ge \cdots
\ge \Lambda_N$.

The next step is to truncate the basis space and retain $D_{cut}$
($\le N$) largest singular values and the corresponding vectors. $M$
is then replaced by an approximate expression
\begin{equation} \label{eq:T1}
M_{ij, kl} \approx \sum_{m= 1}^{D_{cut}} U_{ij , m} \Lambda_m V_{kl
, m} .
\end{equation}
The corresponding truncation error is defined by
\begin{equation}
\varepsilon (\Lambda ) = \frac{\sum_{m>D_{cut}} \Lambda_m
}{\textrm{Tr} \Lambda }
\end{equation}
Eq.~(\ref{eq:T1}) minimizes the truncation error of $M$. However, it
does not consider the influence of the rest of lattice (i.e.
environment) to $M$. In real systems, what needs to be minimized is
acturally the truncation error of the partition function $Z$. This
means that the truncation error is only locally minimized by the
TRG\cite{Levin2007}. For the spin-1/2 Ising model with $D=2$, the
truncation error is generally very small except in the vicinity of
the critical point. However, if the bond degrees of freedom $D$
becomes large, the truncation error increases dramatically. This may
cause a big error in the final result.

\begin{figure}[tbp]
\includegraphics[width=0.6\columnwidth]{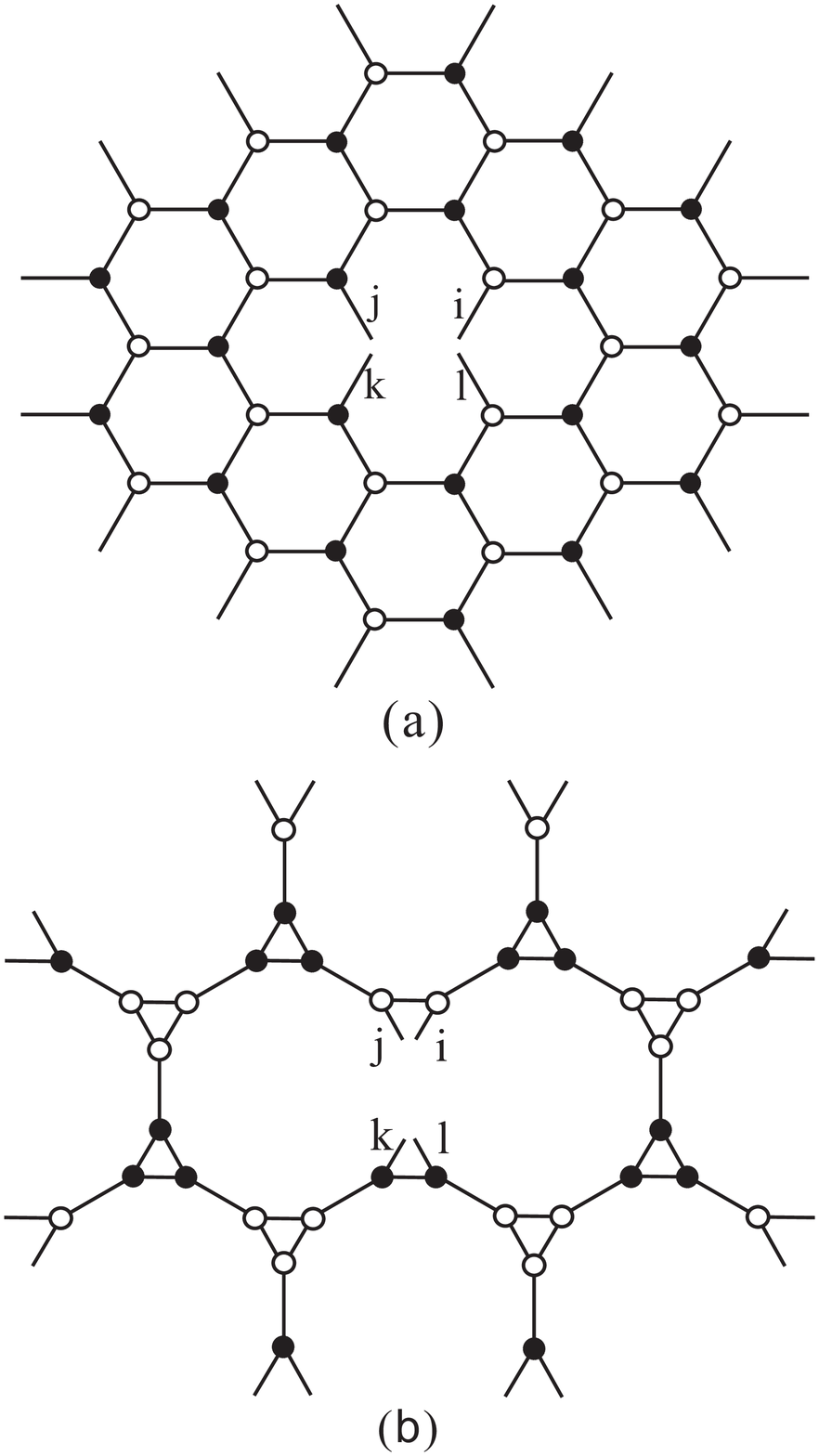}
\caption{Configuration of an environment lattice (a) and that after
one TRG iteration (b). } \label{fig:env}
\end{figure}

To understand this more clearly, let us rewrite the partition
function (\ref{eq:z}) as
\begin{equation}
Z = \textrm{Tr} M M^{e} =  \sum_{ij,kl} M_{ij,kl} M^{e}_{kl,ij},
\end{equation}
where $M^{e}$ is the contribution from the environment lattice
defined in Fig.~\ref{fig:env}. $M^e$ is defined by tracing out all
bond indices in the environment lattice excluding those connecting
with the two vertices on which $M$ is defined. This formula
indicates that to reduce the error in $Z$, one needs to minimize the
truncation error of $MM^e$, rather than that of $M$.

Fig.~\ref{fig:env}(a) shows the configuration of an environment
lattice. In the rewiring and truncation of $M$-matrix, there is no
need to evaluate $M^e$ rigourously. We propose to evaluate $M^e$
iteratively using the TRG method. The configuration of the
environment after one TRG iteration before decimation is shown in
Fig.~\ref{fig:env}(b). By contracting all the internal bonds
connecting small triangles, a decimated environment lattice, whose
configuration is similar to Fig.~\ref{fig:env}(a), is obtained. This
iteration can be repeated until $M^e$ is converged. Generally we
find that the values of $M^e$ such obtained are sufficiently
accurate after 5 to 10 iterations, the corresponding numbers of
environment lattice sites are $2\times 3^5 -2$ and $2 \times
3^{10}-2$, respectively.

\begin{figure}[tb]
\centerline{~
\includegraphics[width=0.9\columnwidth]{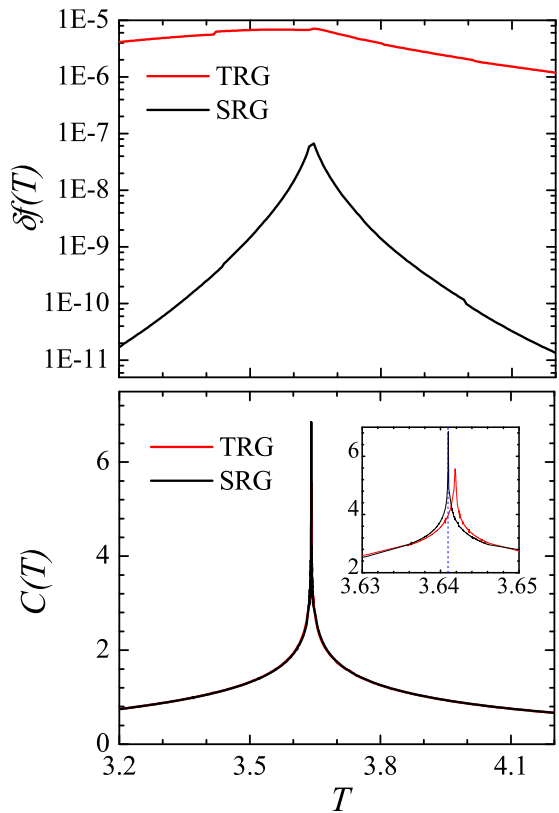} }
\caption{(color online) Comparison of the relative error of the free
energy $\delta f (T)= |f(T) - f_{ex}(T)| / f_{ex}(T)$ (upper panel)
and the specific heat (lower panel) as functions of temperature for
the Ising model on triangular lattices obtained using TRG (red) and
SRG. $f_{ex}(T)$ is the exact result calculated using the formula
given in Ref.~\cite{Wannier1950}. The dotted line in the Inset is
the exact critical temperatures $T_{c} =4/\ln 3$ } \label{fig:CT}
\end{figure}

In the minimization of the truncation error of $M M^e$, it is better
to treat the row $ij$ and column $kl$ indices of $M$ as
symmetrically as possible. To do this, let us first do a singular
value decomposition for $M^e$
\begin{equation}
M^{e} = U_e \Lambda_e V_e^\dagger,
\end{equation}
where $U_e$ and $V_e$ are two unitary matrices and $\Lambda_e$ is a
semi-positive diagonal matrix. Then we can define a new matrix
\begin{equation}
\tilde{M} = \Lambda_e^{1/2} V_e^\dagger M U_e \Lambda_e^{1/2} ,
\label{eq:M2}
\end{equation}
and show that
\begin{equation}
Z = \textrm{Tr} \tilde{M} .
\end{equation}
Thus to minimize the error in $Z$, one needs only to minimize the
truncation error of $\tilde{M}$.

Now let us take a singular value decompostion for $\tilde{M}$
\begin{equation}
\tilde{M} = \tilde{U} \tilde{\Lambda} \tilde{V}^\dagger .
\end{equation}
Again, $\tilde{U}$ and $\tilde{V}$ are two unitary matrices.
$\tilde{\Lambda}$ is a semi-positive diagonal matrix whose diagonal
matrix elements are arranged in descending order. Then we can
truncate the basis space by keeping the $D_{cut}$ largest singular
values of $\tilde{\Lambda}$. By substituting the approximate
$\tilde{M}$ back into Eq.~(\ref{eq:M2}), one can find that
\begin{equation}\label{eq:T3}
M_{ij,kl} \approx  \sum_{n = 1}^{D_{cut}} S^a_{n,ij} S^b_{n,kl} ,
\end{equation}
where
\begin{eqnarray}
S^a &=& \tilde{\Lambda}^{1/2} \tilde{U}^\dagger
\Lambda_e^{-1/2} V_e^\dagger ,\\
S^b &=& \tilde{\Lambda}^{1/2} \tilde{V}^\dagger \Lambda_e^{-1/2}
U_e^\dagger
\end{eqnarray}
are the two tensors defined in the rewired lattice. Finally one can
follow the steps introduced in Ref.~\cite{Levin2007} to update
tensors $T^a$ and $T^b$ in a squeezed lattice by taking the coarse
grain decimation of $S^a$ and $S^b$. This completes a full cycle of
SRG iteration. By repeating this procedure, one can finally obtain
the value of partition function in the thermodynamic limit.

We have applied this SRG method to the spin-1/2 Ising model on
triangular lattices. Fig.~\ref{fig:CT} compares the relative error
of the free energy and the specific heat obtained using the SRG with
those using the TRG. The number of sites is $3^{30}$ and $D_{cut} =
24$. In the SRG calculation of $M^e$, 10 iterations are used. For
the free energy, we find that the SRG can improve the accuracy for
more than five orders of magnitude far away from the critical point,
and more than two orders of magnitude at the critical point. The
critical temperature $T_c$ can be determined from the peak position
of the specific heat. As shown in the Inset of Fig.~\ref{fig:CT},
the value of $T_c$ obtained with SRG is more than two orders of
magnitude more accurate than that obtained with TRG. Furthermore,
from our calculation, we find that the improvement of the SRG over
the TRG becomes more and more pronounced with increasing $D_{cut}$.

It is straightforward to extend the SRG to study ground state
properties of a quantum system with tensor-network wavefunction. The
two-dimensional tensor-network wave function can be accurately
determined using the projection approach we recently
proposed\cite{Jiang2008}. After that one can use the SRG to evaluate
the expectation values of the tensor-network state\cite{note}.

\begin{figure}[tbp]
\hspace{-3mm}\includegraphics[width=1.05\columnwidth]{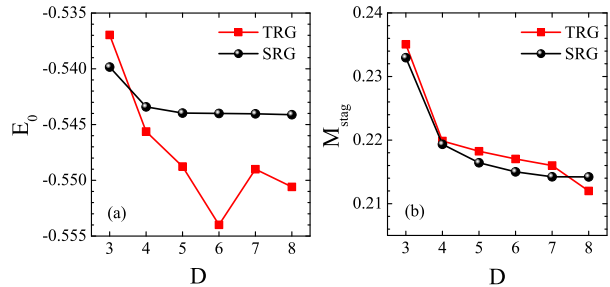}

\caption{ (color online) (a) The ground state energy per site $E_0$
and (b) the staggered magnetization $M_{stag}$ as functions of the
bond degrees of freedom $D$ on honeycomb lattices. }
\label{fig:honey}
\end{figure}

To demonstrate how the SRG can improve the accuracy of the
expectation values of tensor-network states, we have applied the SRG
to the Heisenberg model on honeycomb lattices. The ground state
wavefunction is assumed to have the following tensor-network form
\begin{eqnarray}
|\Psi\rangle &=& \mathrm{Tr} \prod_{i\in b, j\in w } \lambda_{x_i}
\lambda_{y_i} \lambda_{z_i} A_{x_i y_i z_i}
[\sigma_i]  B_{x_j y_j z_i} [\sigma_j] \nonumber \\
&& |\sigma_i \sigma_j \rangle. \label{eq:tns}
\end{eqnarray}
A schematic representation of this tensor-network state on the
honeycomb lattice is shown in Fig.~\ref{fig:z}(b). $\sigma_i$ is the
eigenvalue of spin operator $S_{iz}$. $A_{x_i y_i z_i} [\sigma_i]$
and $B_{x_j y_j z_j} [\sigma_j]$ are the two three-indexed tensors
defined on the black and white sublattices, respectively.
$\lambda_{\alpha_i}$ ($\alpha = x, y, z$) is a positive diagonal
matrix of dimension $D$. The trace is to sum over all spin
configurations and over all bond indices. The tensor corresponding
to $T^a$ in Eq.~(\ref{eq:z}) is now defined by
\begin{eqnarray*}
T^a_{xx^\prime, yy^\prime , zz^\prime} & = & \sum_\sigma
(\lambda_x\lambda_y\lambda_z)^{1/2} A_{xyz}[\sigma] \\
&& A_{x^\prime y^\prime z^\prime} [\sigma ] (\lambda_{x^\prime}
\lambda_{y^\prime} \lambda_{z^\prime})^{1/2} .
\end{eqnarray*}
The bond dimension of this tensor is $D^2$.

Fig.~\ref{fig:honey} compares the SRG with the TRG results for the
ground state energy and the staggered magnetization for the
Heisenberg model on the honeycomb lattice. The number of lattice
sites is $2\times 3^{18}$. The truncation error in the SRG
calculation is less than $\varepsilon_0 \sim 10^{-3}$ and $D_{cut}=
D^2$.  We have used the second order Trotter-Suzuki decomposition
formula to improve the accuracy in the calculation of the ground
state wavefunctions using the projection approach introduced in
Ref.~\cite{Jiang2008}. The staggered magnetization is evaluated
directly from the expectation value of the staggered spin operator
in the ground state in the limit the external staggered magnetic
field approaching zero. This avoids the error in the determination
of the staggered magnetization from the numerical derivative of the
ground state energy at finite staggered magnetic field, as was done
in Ref.~\cite{Jiang2008}. Unlike the TRG results, we find that the
SRG results vary monotonically with $D$ and tend to converge quickly
to the infinite $D$ limit.

For $D=8$, the SRG results of the ground energy and the staggered
magnetization per site are respectively -0.5445 and 0.2142,
consistent with the results obtained by other
methods\cite{Zheng1991,Oitmaa1992,Reger1989}. The accuracy of these
results are still not comparable with those obtained by the
DMRG\cite{White07} and the quantum Monte Carlo method\cite{Sandvik}.
By considering the symmetry of the Hamiltonian, the tensor-network
states with a bond dimension as large as $D\sim 20$ can in principle
be handled. In that case, the SRG results will be further improved.

In conclusion, we have introduced a SRG method to improve
significantly the accuracy in the TRG calculation. This method
differs from the TRG by taking into account the renormalization
effect of environment to the $M$-matrix, similar as the DMRG
contrasting the conventional block renormalization group method. For
the classical Ising model, the relative error of the free energy as
well as other quantities is reduced by more than two to five orders
of magnitude when $D_{cut}=24$ and can be further reduced by
increasing $D_{cut}$, in comparison with the TRG. The SRG, in
combined with the projection method introduced in
Ref.~\cite{Jiang2008}, provides an accurate and efficient tool for
exploring tensor-network ground states of quantum lattice models. It
will play a more and more important role in the study of highly
correlated systems. The physical idea present in this work can be
also generalized to apply to other physical problems where the
system can be divided into two parts and the interplay between them
is important. In particular, if one wants to generalize the
projection method proposed in Ref.~\cite{Jiang2008} to evaluate
time-dependent or thermodynamic quantities, then the SRG correction
to the wavefunction should be considered to minimize the accumulated
Trotter and truncation errors in the iteration.

This work was supported by the NSF-China and the National Program
for Basic Research of MOST, China.

\end{document}